\documentstyle[aps,preprint]{revtex}
\begin{document}
\draft
\preprint{}
\title{Induced On-shell Supersymmetry in Eikonal Scattering}
\author{Parthasarathi Majumdar}
\address{The Institute of Mathematical Sciences, \\CIT Campus, Madras 600 113,
India.}
\maketitle
\begin{abstract}
Generically coupled neutral scalar bosons and chiral fermions are shown,
in the eikonal kinematical limit, to be described by
a reduced (free field) theory  with $N=1$ {\it on-shell} supersymmetry. {\it
Charged}
scalars and spinors turn out to
be described in the eikonal limit by a reduced interacting theory with a
modified
and restricted on-shell $N=1$ supersymmetry. Consequences of such a symmetry
for the
nontrivial scattering amplitudes in this latter case are discussed.
\end{abstract}
\newpage

Spacetime supersymmetry has two major uses in high energy theory : as a means
of
resolving the problems of naturalness and the stability of the gauge
hierarchy in the Standard Model \cite{maj}; as a device
to eliminate spacetime tachyons from the spectrum of string theories
\cite{gerv}. Yet,
there is so far no compelling experimental evidence at all of this symmetry,
even as an
approximate fundamental symmetry of nature.\footnote{The
unification of gauge coupling constants inferred from LEP data can be neither
uniquely attributed to, nor be taken as unambiguous evidence of, spacetime
supersymmetry.} Nor are we
any closer to finding a non-perturbative mechanism for supersymmetry breaking
which
would still lead to a {\it naturally} small cosmological constant for the real
world.
Recent advances in formal aspects of exactly supersymmetric field and string
theories, related to electric-magnetic duality, have so far only a tenuous
connection with reality. While prospects of a headway into strong coupling
situations appear to be good, the dependence on unbroken supersymmetry seems
crucial in the more interesting cases. Despite the unmistakeable beauty of the
mathematical structures it embodies, spacetime supersymmetry remains an enigma.

Is it conceivable, however, that in nature supersymmetry is dynamically {\it
induced},
rather than fundamental? In this letter we point out a possibility as to how a
version of supersymmetry,
realized as a transformation between physical scalar and spinor fields (with
generic
nonsupersymmetric Yukawa and scalar self couplings),
must arise in a certain kinematical limit. Two major characteristics
distinguish
this version of supersymmetry from standard spacetime supersymmetry : first of
all, it is an inevitable consequence of the  kinematical restrictions imposed
on
a system whose parameters (masses and couplings) exhibit no intrinsic feature
of spacetime
supersymmetry; secondly, the transformation laws affect only matter fields
leaving gauge
bosons (including gravitons) inert by assumption. Consequently, this sort of
supersymmetry has little to do with
naturalness, the gauge hierarchy problem or tachyons. However, we shall argue
that
it will have non-trivial dynamical consequences, especially pertaining to small
angle electromagnetic, and possibly gravitational, scattering of bosons and
fermions.

The kinematical limit in question is the so-called eikonal limit -- the limit
of
arbitrarily large
ratio of the squared center-of-mass energy ($s$) to the squared momentum
transfer
($t$). Clearly, these are the kinematics of almost forward scattering with tiny
scattering angles, well-studied in the context of electromagnetism \cite{jac}
and also
including gravitation \cite{thf}-\cite{dm}, for scalar particles. In this
rather
singular limit, transverse photon/graviton exchange between scattering
particles is
severely suppressed; the amplitude is dominated by a semiclassical process
induced
by instantaneous classical {\it shock wave} gauge configurations, and becomes
exactly computable. Corrections to the semiclassical approximation due to
fluctuations around the shock wave become important only if one recedes from
the eikonal limit. Now, within the kinematical restrictions of the eikonal, the
large ratio of the transverse and longitudinal momentum scales allows a scaling
of the longitudinal (lightcone) coordinates, relative to the
transverse ones. In other words, one can transform to a very large momentum
frame,
and use the resulting scaling on fields to determine which of these participate
in the scattering process \cite{ver}, and also the relevant interactions that
survive
such scaling. The resulting reduced theory describes particle scattering in the
eikonal limit {\it exactly}, i.e., without further approximation. Other
features of
the reduced theory include the appearance of certain global symmetries absent
in the
original Lagrangian. These latter aspects will be our major concern in this
paper.

As the simplest example of the appearance of such symmetries, consider a real
scalar
field with the standard action in Minkowski 4-space
\begin{equation}
S~=~\int d^4x~\left[ \frac12 (\partial_{\mu} \phi)^2 ~-~V(\phi) \right] ~,
\label{rescl}
\end{equation}
where, the potential $V(\phi)$ might have a mass term $\frac12 \mu^2 \phi^2$
and higher
order self-coupling terms. Following \cite{ver}, we perform the following
scaling of the lightcone coordinates : $x^{\pm}~\rightarrow~\xi x^{\pm}$, with
$x^{\pm} \equiv x^0 \pm x^3$, and $\xi \sim t/s$. The transverse coordinates
${\vec x}_{\perp}$
remained unchanged under the scaling. Under this scaling, of course, the scalar
field undergoes no change, while the lightcone derivatives do scale by
$\xi^{-1}$. Taking
account of the scaling of the integration measure, the net effect of these on
the action is
\begin{equation}
S~\rightarrow~ \int d^4 x ~ \partial_+ \phi \partial_- \phi ~+~ O ( \xi^2)~.
\label{sceik}
\end{equation}
In the eikonal limit $\xi \rightarrow 0$, the action reduces to a free field
action with
only lightcone derivatives; apart from the obvious symmetry associated with the
conservation of the number of $\phi$ particles, this reduced theory is also
invariant
under $\phi \rightarrow \phi + constant$, a symmetry absent in the original
formulation.
Of course the price to pay is the loss of manifest Lorentz invariance. In any
event,
it is clear that $\phi$-particles scatter in the forward direction with unit
amplitude
because of the decoupling seen above. This result is consistent with earlier
assertions \cite{jac} that for scalar exchanges, the eikonal approximation is
never
dominant; the theory is rendered trivial in this approximation.

Generalization to a theory of self-interacting complex scalars is
straightforward : the
reduced action in the eikonal limit becomes
\begin{equation}
S_{red}~=~\int d^4x ~ \partial_{(+} \phi^* \partial_{-)} \phi~
\label{csceik} \end{equation}
which continues to be invariant under global rotations $\phi \rightarrow
e^{i \theta} \phi$, just as the original action (assuming that the
potential is a function only of $|\phi|$). In addition, the number of
positive and negative charges are separately conserved, and the reduced
action is also invariant under the shift of $\phi$ by a constant. Thus,
in this limit, the theory has a two (real) dimensional space of vacua;
one of the flat directions can be identified as a Goldstone direction if the
potential
$V(|\phi|)$ of the theory before scaling has a minimum away from the
origin, and the other (the Higgs) is a {\it modulus}. We shall return to
this reduced theory later.

A similar symmetry enhancement takes place with the free massive Dirac
theory in four dimensions as well;
\begin{equation}
S~=~\int d^4 x~ {\bar \psi} \left ( i \gamma^{\mu} \partial_{\mu} ~-~ m \right
) \psi~.
\label{spin} \end{equation}
Under the above scaling of the lightcone coordinates, the spinor field $\psi
\rightarrow
\xi^{-\frac12} \psi$, so that
\begin{equation}
S~\rightarrow~ \int d^4x ~ i {\bar \psi} \gamma_{(+} \partial_{-)} \psi ~+~
O(\xi)~,
\label{speik}
\end{equation}
implying that in the eikonal limit $\xi \rightarrow 0$ one is left with a {\it
massless}
theory
with only lightcone derivatives, and the (global) chiral symmetry under $\psi
\rightarrow
e^{i \gamma_5 \theta} \psi$. Of course, for the free theory the symmetry simply
ensures the
separate conservation of left handed and right handed fermion number.

Now consider the most general theory of a Dirac spinor and a complex
scalar field, with canonical kinetic energy terms,
\begin{equation}
S~=~\int d^4x \left [ |\partial_{\mu} \phi|^2 ~-~V(|\phi|) ~+~{\bar \psi}
\left (i \gamma^{\mu} \partial_{\mu}~-~H(\phi) \right) \psi \right]~.
\label{scsp} \end{equation}
The functions $V(\phi)$ and
$H(\phi)$ are assumed to be completely arbitrary and mutually independent,
unlike in
an intrinsically supersymmetric theory where they are both derived from a
holomorphic
superpotential.

Under the scaling of the scalar and spinor fields induced by the scaling of the
lightcone
coordinates in the passage to the large momentum frame, this action
reduces to
\begin{equation}
S_{red}~=~\int d^4x~\left[ \partial_{(+} \phi^* \partial_{-)} \phi ~+~{\bar
\psi} \gamma_{(+} \partial_{-)} \psi \right]~, \label{ssfre}
\end{equation}
which is once again a free field theory action. However, $S_{red}$ has an
additional
symmetry property: the Lagrangian density changes by a total lightcone
derivative
\begin{equation}
\delta {\cal L}_{red}~=~\partial_{(+} \left ({\bar \epsilon} \partial_{-)}
\psi~\phi
\right) ~\label{sstrns} \end{equation}
under the transformations
\begin{equation}
\delta \phi ~=~ {\bar \epsilon} \psi~~,~~ \delta \psi~=~-i\gamma_{(+}
\partial_{-)} \phi \epsilon~~, \label{sstr0} \end{equation}
where $\epsilon$ is a spacetime-independent spinorial parameter. While
reminiscent of
standard spacetime supersymmetry, the glaring absence of any auxiliary field
reminds
us of the {\it on-shell} nature of this version
\cite{gate}. Furthermore, irrespective of its symmetries, a free theory is
intrinsically
of extremely limited interest.

We now explore the possibility of realizing such a symmetry with less trivial
consequences. To this end, we endow the fields with electric charge and couple
them
minimally to electromagnetism. The behaviour of the Maxwell action under the
scaling of
lightcone coordinates is well-known \cite{ver}, \cite{jac}
\begin{equation}
S_{Max}~=~\int d^4 x~\left[ \frac{1}{\xi^2} F_{+-}^2~+~F_{\alpha
i}^2~+~O(\xi^2) \right]~ ,
\label{maxsc} \end{equation}
where, $\alpha=\pm$. Now, in the eikonal limit $\xi \rightarrow 0$ the first
term
explodes, so that the dominant contribution to the partition function comes
from
gauge field configurations for which $F_{+-}~=~0$ which implies that
$A_{\pm}~=~\partial_{\pm} \Omega$. Thus, only the second term survives in the
eikonal
limit. The matter action, modified to include the coupling to electromagnetism,
reduces
in this limit to
\begin{equation}
S_{red}~=~S_0~+~S_{int}~,\label{mactn} \end{equation}
where, $S_0$ is the reduced action given in eq. (\ref{ssfre}) and
\begin{equation}
S_{int}~=~\int d^4 x~\left [ ie A_{(+} \left ( \phi^* \partial_{-)} \phi -c.c.
- {\bar
\psi} \gamma_{-)} \psi \right ) ~+~ e^2 A_+A_- |\phi|^2 \right]~.
\label{inter} \end{equation}
All interactions of the transverse gauge potentials $A_i, i=1,2$ drop out in
the $\xi
\rightarrow 0$ limit, thereby rendering it a free field not warranting further
consideration. We should point out a crucial assumption : both scalar and
spinor fields
have the same electric charge $e$. This is imperative for the symmetry
considerations
to follow. Note, however, that, because of $S_{int}$ the particles no longer
forward
scatter with unit amplitude in the eikonal domain.

Because we prefer not to introduce new fields like the gaugino, we must assume
that
whatever symmetry transformations the matter fields are subjected to must leave
the
gauge fields invariant. These assumptions
suffice to demonstrate that, after some arithmetic, under the transformations
\begin{equation}
\delta \phi~=~{\bar \epsilon} \psi~~;~~\delta \psi~=~i \gamma_{(+} D_{-)} \phi
\epsilon~~\label{sstrns} \end{equation}
the Lagrangian density in $S_{red}$ transforms into
\begin{equation}
\delta {\cal L}_{red}~~\sim~~ \partial_{(+}\left [~{\bar \epsilon} ( D_{-)}
\psi)
\phi^* ~+~c.c.\right]~~, \label{ss} \end{equation}
where $D_{\pm}$ is the $U(1)$ covariant lightcone derivative
$$D_{\pm} \phi~~ \equiv~~ \partial_{\pm} \phi~-~ie A_{\pm} \phi ~. $$
In arriving at the result in eq. (\ref{ss}), we have made repeated use of the
commutator $[~D_+~,~D_-~]~\sim~ F_{+-}~=~0$ on all fields. Observe that the
transformation of the fermion fields is non-linear, involving both the gauge
potential
and the scalar. This is yet again a departure from standard
super-transformation laws.
Thus, it is not possible to interpret the transformations as being the
`square root' of spacetime translations. The question of interest is, however,
whether this symmetry imposes non-trivial restrictions on boson and fermion
scattering
amplitudes, in analogy
with restrictions on helicity-flip amplitudes obtained by Grisaru and Pendleton
\cite{gri}
more than two decades ago within standard globally supersymmetric 4d field
theories.
Such restrictions, if any, would be all the more predictive here
given that the original theory (\ref{scsp}) has no supersymmetry to begin with.

The simplest way to visualize the scattering processes we have in mind is to
use the
electromagnetic shock wave description \cite{thf}, \cite{ver}, \cite{jac}.
Thus, for
two-particle scattering, one chooses a Lorentz frame in which one of the
particles is
moving almost luminally, carrying with it a plane-fronted electromagnetic {\it
shock} wave
with the (infinitely extended) shock plane transverse to the direction of
propagation. The
fields due to this particle vanish everywhere except on the plane where they
have a $\delta$
function singularity. Consequently, a test particle, slow-moving relative to
our chosen
frame, experiences no force except when the shock front passes it. The
resulting phase
factor induced instantaneously by the shock wave in the wave function of the
test particle
manifests in a non-trivial scattering amplitude (calculated as an overlap of
the test
particle wave functions before and after the impact \cite{thf}). In the
center-of-mass
frame of the particles, the amplitude can be calculated directly by a path
integral approach
\cite{ver}, \cite{jac}, and amounts to determining the amplitude for the
elastic collision
of two shock wave fronts in the eikonal kinematics. The issue of our concern
here
is whether the scattering amplitudes for elastic eikonal collisions of bosons
are
related to those of fermions.

To this end it suffices to restrict the matter field theory (\ref{mactn}) to
on-shell
field configurations, i.e., solutions to the free equations of motion obtained
from varying the action $S_0$; it also makes sense to invoke the {\it
classical}
eikonal approximation, also referred to as the `geometrical optics'
approximation
\cite{lan}. In this approximation, the fields, expressed as complex-valued
functions, have
phases which vary extremely rapidly relative to the variation of the moduli.
E.g., the
complex scalar field $\phi$ has the polar decomposition
$\phi(x)~=~\rho(x)~e^{i\theta}(x)$;
in the eikonal approximation, we make the approximation $\partial_{\pm}\rho ~
\ll~ \rho~
\partial_{\pm} \theta$, so that the modulus $\rho~=~\rho({\vec x}_{\perp})$.
Thus,
\begin{equation}
S_0~\approx~\int d^4 x~ \left [ ~\rho^2~\partial_+ \theta \partial_- \theta~+~i
{\bar \psi}
\gamma_{(+} \partial_{-)} \psi ~\right]~~\label{ff} \end{equation}
The corresponding equations of motions are
\begin{eqnarray}
\rho^2 ~\partial_+ \partial_- \theta ~&=&~ 0~ \label{eoma} \\
\rho ~\partial_+ \theta \partial_- \theta ~&=&~0~ \label{eomb} \\
\partial_-~[~\gamma_- \gamma_+ \psi~]~&=&~0~=~\partial_+~[~\gamma_+ \gamma_-
\psi~]~. \label{eomc} \end{eqnarray}
Since $\rho~\neq~0$, eq. (\ref{eoma}) has the solution
\begin{equation}
\theta(x^{\pm}, {\vec x}_{\perp})~=~\theta^{(+)}~(x^+,{\vec
x}_{\perp})~+~\theta^{(-)}~(x^-,{\vec x}_{\perp}) ~.\label{scsol}
\end{equation}
Eq. (\ref{eomb}) is simply the on-shell constraint for the (massless) scalar
particles.
As for eq. (\ref{eomc}), defining $\chi^{(\pm)}~\equiv~\gamma^{\pm}
\gamma^{\mp} \psi$, one has the solutions
\begin{equation}
\chi^{(+)}~=~\chi^{(+)}~(x^+, \vec{x}_{\perp})~~,~~\chi^{(-)}~=~\chi^{(-)}(x^-,
{\vec
x}_{\perp})~. \label{spsol} \end{equation}

Using these solutions, the interacting part of the action assumes the form
\begin{eqnarray}
S_{int}~&=&~e~\int~\left [~A_-~\left(~i\rho^2 \partial_+ \theta^{(+)} ~-~{\bar
\chi}^{(+)} \gamma_+ \chi^{(+)}~\right) \right] \nonumber \\
{}~&+&~ e~\int ~\left[~A_+~\left(~-~i \rho^2 \partial_- \theta^{(-)}~-~{\bar
\chi}^{(-)}
\gamma_- \chi^{(-)}~\right) \right] \nonumber \\
{}~&+&~ e^2 \int \rho^2~A_+~A_-~~, \label{intera} \end{eqnarray}
where $\int~\equiv~\int d^4x$. Thus, the first two lines represent the usual $j
\cdot A$
type of gauge interaction, with
\begin{equation}
j_{\pm}=j_{B,\pm}+j_{F,\pm}~, \end{equation}
where,
\begin{eqnarray}
j_{B,\pm}~&=&~\pm i~\rho^2~\partial_{\pm}\theta^{(\pm)}~\equiv~
\partial_{\pm}k_{B}^{(\pm)}(x^{\pm}, {\vec x}_{\pm})~\nonumber \\
j_{F,\pm}~&=&~-{\bar \chi}^{(\pm)} \gamma_{\pm}
\chi^{(\pm)}~\equiv~\partial_{\pm}
k_F^{(\pm)}~. \label{curnts}
\end{eqnarray}
The first line of eq. (\ref{curnts}) defines $k_B$ in the same manner as in
ref.
\cite{jac}. For the spinor current, we have appealed to `bosonization' of the
fermionic
field considered as functions only on the null plane; this does not entail any
loss of
generality since in the kinematical region of interest, derivatives with
respect to the
transverse coordinates do not appear in the interaction Lagrangian. thus
$k_F^{(\pm)}(x^{\pm}, {\vec x}_{\perp})$ is a Lorentz scalar.  Both currents of
course are conserved via eq.s (\ref{scsol}) and (\ref{spsol}).

Observe also that the
`transverse' components $A_i$ have
decoupled from matter in our kinematic regime, and can therefore be set to zero
without
loss of generality. Recall now that $A_{\pm}=\partial_{\pm}
\Omega$ as a consequence of the constraint $F_{+-}=0$. Thus, imposing the
Lorentz-Landau
gauge condition \cite{jac} on the gauge potential implies
\begin{equation}
\partial_+~\partial_-~\Omega~~=~~0~\label{hrmnc} \end{equation}
so that, $\Omega~=~\Omega^{(+)}(x^+, {\vec x}_{\perp})~+~\Omega^{(-)}(x^-,
{\vec
x}_{\perp})$. With these simplifications, both the Maxwell and the interaction
Lagrangians can be expressed as total derivatives on the null plane :
\begin{equation}
S_{Max}~=~\int~ \left[~\partial_-\left (~\Omega^{(-)}~ \nabla_{\perp}^2~
\partial_+ \Omega^{(+)}~\right)~+~ \left(~+~\leftrightarrow~-~\right)~\right]~
\label{max}
\end{equation}
\begin{eqnarray}
S_{int}~&=&~ e~\int~\left[~\partial_-\left \{\left(~j_{B,+}~+~j_{F,
+}~\right)~\Omega^{(-)}~\right \}~+~\left(~+~\leftrightarrow~-~\right)~\right]
\nonumber \\
{}~&+&~ \frac12
e^2~\int~\left[~\partial_+~\left(~\rho^2~\Omega^{(+)}~\partial_-
\Omega^{(-)}~\right) ~+~\left(~ +~\leftrightarrow~-~\right)~\right]~ .
\label{actns}
\end{eqnarray}
The action $S~=~S_{Max}~+~S_{int}$ thus reduces to a field theory
`living' on the three dimensional space composed by the transverse plane and
the
boundary of the null plane. Parametrising the latter by $\tau$ and indicating
the $\tau$-derivative by an overdot, eq. (\ref{actns}) can be recast into
\cite{jac}
\begin{equation}
S_{Max}~=~\int d^2{\vec x}_{\perp} \oint d\tau ~\left[ ~{\bar \Omega}^{[(-)}
\nabla_{\perp}^2 {\dot {\bar \Omega}^{(+)]}} \right]~,\label{mxred}
\end{equation}
where, the bar on $\Omega$ indicates that it is evaluated on the boundary of
the null
plane. Similarly,
\begin{eqnarray}
S_{int}~&=&~\int d^2{\vec x}_{\perp} \oint d\tau~\left [~\left(~{\dot
k}_B^{(+)}~+~{\dot
k}_F^{(+)}~\right)~{\bar
\Omega}^{(-)}~+~\left(~+~\leftrightarrow~-~\right)~\right]~\nonumber \\
{}~&+&~e^2~\int d^2{\vec x}_{\perp} \oint d\tau~\partial_{\tau}~\left(~\rho^2
{\bar
\Omega^{(+)}} {\bar \Omega^{(-)}}~\right)~~\label{intrd}
\end{eqnarray}
The seagull term drops out upon integration over $\tau$, assuming that the
gauge
degrees of freedom are single-valued on the boundary of the null plane. Note
that, in
addition to the well-known interaction of the bosonic current \cite{jac}, the
action
above includes the fermionic current. Thus, scattering amplitudes may be
computed from
the expectation values of the three vertex operators
\begin{eqnarray}
V_{BB}~&=&~\exp~\left[~i~\oint d\tau \int d^2{\vec x}_{\perp}~\left (~{\dot
k}_B^{(+)} {\bar
\Omega}^{(-)}~+~{\dot k}_B^{(-)} {\bar \Omega}^{(+)}~\right)~ \right] \nonumber
\\
V_{FF}~&=&~\exp~\left[~i~\oint d\tau \int d^2{\vec x}_{\perp}~\left(~{\dot
k}_F^{(+)} {\bar
\Omega}^{(-)}~+~{\dot k}_F^{(-)} {\bar \Omega}^{(+)}~\right)~\right] \nonumber
\\
V_{BF}~&=&~\exp~\left[~i~\oint d\tau \int d^2{\vec x}_{\perp}~\left(~{\dot
k}_B^{(+)} {\bar
\Omega}^{(-)}~+~{\dot k}_F^{(-)} {\bar \Omega}^{(+)}~\right)~ \right]~.
\label{verop}
\end{eqnarray}

The essential difference between the vertex operators in eq. (\ref{verop}) stem
from the
difference between $k_B$ and $k_F$. Indeed, as functions of the basic fields
they are quite
distinct. However, the distinction blurs when two point-particle scattering in
the
eikonal limit is considered. The main reason for this, in the shock wave
picture, has
to do with the fact that, the restriction of the electric and magnetic fields
to the
transverse shock plane implies that helicity-flip effects are absent in the
eikonal
regime. The shock wave impinging upon the test particle does not `see' its
spin. Another way of seeing this is to appeal to the so-called Gordon
decomposition of
the spinor current and observe that the Pauli term cannot contribute in the
eikonal
approximation. The Dirac term, on the other hand, for almost luminal particles
will
reduce to a form which is almost identical to the current for an
ultrarelativistic point boson. We should point out that these somewhat
heuristic
considerations hold only for point particles; their validity beyond that is not
claimed.
In any event, the net upshot is that the amplitudes
\begin{eqnarray}
<~V_{BB}~>~&=&~<~V_{FF}~>~=~<~V_{BF}~>~\nonumber\\
{}~&=&~ \exp\left[\oint d\tau \int d^2 {\vec x}_{\perp} d^2 {\vec x}_{\perp}'
\left(\log|{\vec x}_{\perp}-{\vec x}_{\perp}'|
k^{(+)}(\tau, {\vec x}_{\perp})~{\dot
k}^{(-)}(\tau, {\vec x}_{\perp}') + (+ \leftrightarrow -) \right) \right]
\label{ampl}
\end{eqnarray}
where,
\begin{equation}
k^{(\pm)}(x^{\pm}(\tau), {\vec
x}_{\perp})~\equiv~e~\Theta~\left(~x^{\pm}(\tau)~-~x^{(A)\pm}~\right)~\delta^{(2)} ({\vec
x}_{\perp}~-~{\vec x}^{(A)})~ \label{kay} \end{equation}
with $A=1,2$ for two-particle scattering, and $\Theta$ is the unit step
function. This
result is the same as in ref. \cite{jac} for scalar particles.

The identity of the bosonic and fermionic amplitudes in eq. (\ref{ampl}) has
rather
remarkable ramifications. First of all, helicity-flip amplitudes vanish because
of the
behaviour of electromagnetism in the eikonal domain, consistent with the
results of
\cite{gri}, as anticipated on the basis of the on-shell supersymmetry of the
reduced
action. Secondly, the fact that the Yukawa and scalar self-couplings drop out
in the
eikonal kinematics might have implications for {\it pion-nucleon} dynamics.
Consider,
e.g., the
Gell Mann-Levy $\sigma$ model \cite{gel}, proposed decades ago as a model for
PCAC and low energy theorems in pion-nucleon interactions. It is amusing to
examine
this model in the eikonal regime for pion-nucleon, pion-pion and proton-proton
elastic
scattering processes.  The action
has a chiral $SU(2) \times SU(2)$ symmetry in the absence of nucleon masses; it
is given by
\begin{eqnarray}
S_{\sigma}~&=&~\int~\left (~ {\bar N} i \gamma^{\mu} \partial_{\mu}
N~+~\frac12 (\partial_{\mu} \sigma)^2~+~\frac12 (\partial_{\mu} {\vec
\pi})^2~\right)
\nonumber \\ ~&-&~ \int~\left(~\lambda~(~\sigma^2 ~+~{\vec
\pi}^2-a^2~)^2~-~h~{\bar
N}(~\sigma~+~i~\gamma_5~{\vec \tau} \cdot {\vec \pi}~)N~ \right)~,
\label{sigma}
\end{eqnarray}
where, $N$ is the nucleon isodoublet, ${\vec \pi}$ is the isotriplet of pions.
The isospin
symmetry is, of course, broken by the electromagnetic interactions which must
be there
between protons and charged pions, but this breaking is a tiny perturbation on
the strong interactions that usually dominate pion-nucleon dynamics. However,
our
considerations above would imply that, in the eikonal regime,
strong interactions, as depicted in the $\sigma$ model,  would effectively be
suppressed, so that, in
this kinematical region, electromagnetism should take over. Furthermore,
small-angle
elastic scattering of charged pions should have identical amplitudes as those
of
protons. In other words, their behaviour in these kinematics should be
extremely
similar, to the extent that it is describable in terms of the $\sigma$ model.
Perhaps
an analysis of the data for elastic $p-p$, $\pi-\pi$ and $\pi-p$ scattering at
$\sqrt{s} \gg 1Gev$ and $t\rightarrow 0$ (almost-forward scattering) is in
order to
test the validity of these `predictions'.

Finally, the induced supersymmetry discerned above in eikonal scattering
through
electromagnetism might reappear for gravitational scattering of light point
particles
in Minkowski space. If so, it ought to find application in the analysis of
Hawking
radiation from black holes, taking into account the interaction of the outgoing
radiation with collapsing matter, recalling that \cite{thf2}, \cite{ver2} such
interactions typically involve large centre-of-mass momenta and small momentum
transfers. The intriguing question to probe in this problem is how the
supersymmetry
disappears from the Hawking spectrum.

We thank S. Das, A. Dasgupta and R. Kaul for useful discussions.


\begin{references}
\bibitem{maj} R. Kaul and P. Majumdar, Bangalore preprint (1981), unpublished;
E.
Witten, Nucl. Phys. {\bf B188} (1981) 513; S. Dimopoulos and H. Georgi, Nucl.
Phys.
{\bf B193} (1981) 150; N. Sakai, Z. Phys. {\bf 11} (1981) 153; R. Kaul and P.
Majumdar, Nucl. Phys. {\bf B199} (1982) 36; R. Kaul, Phys. Lett. {\bf 109B}
(1982) 19.
\bibitem{gerv} J.-L. Gervais and B. Sakita, Nucl. Phys. {\bf B34} (1971) 62.
\bibitem{jac} R. Jackiw, D. Kabat and M. Ortiz, Phys. Lett. {\bf B277} (1992)
148.
\bibitem{thf} G. 't Hooft, Phys. Lett. {198B} (1987) 61, Nucl. Phys. {\bf B304}
(1988)
867.
\bibitem{amat} D. Amati, M. Ciafaloni and G. Veneziano, Nucl. Phys. {\bf B347}
(1990)
550 and references therein.
\bibitem{lous} D. Loust\'o and N. S\'anchez, Int. Jour. Mod. Phys., {\bf A5}
(1990)
915.
\bibitem{ver} H. Verlinde and E. Verlinde, Nucl. Phys. {\bf 371} (1992) 246;
Princeton
University preprint PUPT-1319 (1993), unpublished.
\bibitem{kab} D. Kabat and M. Ortiz, Nucl. Phys. {\bf 388} (1992) 148.
\bibitem{dm} S. Das and P. Majumdar, Phys. Rev. Lett. {\bf 71} (1994) 2524;
Phys.
Rev. {\bf D51} (1995) 5664; Phys. Lett. {\bf 348B} (1995) 349. For a review
see, P.
Majumdar, IMSc preprint IMSc/94-95/64, hepth 9503206, to appear in the
Proceedings of
the International Conference on the Physics of the Planck scale, Puri, India,
December
1994.
\bibitem{gate} S. J. Gates, Jr., M. Grisaru, M. Rocek and W. Siegel, {\it
Superspace}
(Benjamin-Cummings Inc., 1983).
\bibitem{gri} M. Grisaru and H. Pendleton, Nucl. Phys. {\bf B124} (1977) 81.
See also,
T. Hagiwara and P. Majumdar, Nucl. Phys. {\bf B191} (1981) 170.
\bibitem{lan} L. Landau and E. Lifshitz, {\it The Classical Theory of Fields},
pp.
129-132 (Pergamon, 1975).
\bibitem{gel} M. Gell-Mann and M. Levy, Nuov. Cim. {\bf 16} (1960) 705.
\bibitem{thf2} G. 't Hooft, Nucl. Phys. {\bf B335} (1990) 136.
\bibitem{ver2} H. Verlinde and E. Verlinde, CERN-Princeton preprint
CERN-TH.7469/94,
PUPT-1504 (1995).

\end{references}
\end{document}